\documentclass[conference]{IEEEtran}
\IEEEoverridecommandlockouts

\def \toolname {\textsc{Asdf} }

\usepackage{amsmath,amssymb,amsfonts}
\usepackage{array}
\usepackage{algorithmic}
\usepackage{graphicx}
\usepackage{textcomp}
\usepackage{xcolor}
\usepackage{float}
\usepackage{booktabs}
\usepackage{tabularx}
\usepackage{adjustbox}
\usepackage[many]{tcolorbox}
\usepackage{balance}
\usepackage{hyperref}
\usepackage{tipa}
\usepackage{svg}
\usepackage{color, colortbl}
\def\BibTeX{{\rm B\kern-.05em{\sc i\kern-.025em b}\kern-.08em
    T\kern-.1667em\lower.7ex\hbox{E}\kern-.125emX}}
\begin{document}

\title{ASDF: A Differential Testing Framework for Automatic Speech Recognition Systems}

\newcommand{\linebreakand}{%
  \end{@IEEEauthorhalign}
  \hfill\mbox{}\par
  \mbox{}\hfill\begin{@IEEEauthorhalign}
}


\author{
    \IEEEauthorblockN{Daniel Hao Xian Yuen,\IEEEauthorrefmark{1} Andrew Yong Chen Pang,\IEEEauthorrefmark{1} Zhou Yang,\IEEEauthorrefmark{2} Chun Yong Chong,\IEEEauthorrefmark{1} Mei Kuan Lim,\IEEEauthorrefmark{1}} David Lo\IEEEauthorrefmark{2}
    \IEEEauthorblockA{\IEEEauthorrefmark{1}School of Information Technology, Monash University Malaysia
    \\\{dyue0003, apan0027, chong.chunyong, lim.meikuan\}@monash.edu}
    \IEEEauthorblockA{\IEEEauthorrefmark{2}School of Computing and Information Systems, Singapore Management University
    \\\{zyang, davidlo\}@smu.edu.sg
}
}
\maketitle

\begin{abstract}
Recent years have witnessed wider adoption of Automated Speech Recognition (ASR) techniques in various domains.
Consequently, evaluating and enhancing the quality of ASR systems is of great importance.
This paper proposes \textsc{Asdf}, an \underline{A}utomated \underline{S}peech Recognition \underline{D}ifferential Testing \underline{F}ramework for to test ASR systems. 
\toolname extends an existing ASR testing tool, the CrossASR++, which synthesizes test cases from a text corpus.
However, CrossASR++ fails to make use of the text corpus efficiently and provides limited information on how the failed test cases can improve ASR systems. 
To address these limitations, our tool incorporates two novel features: (1) a \textit{text transformation module} to boost the number of generated test cases and uncover more errors in ASR systems and (2) a \textit{phonetic analysis module} to identify on which phonemes the ASR system tend to produce errors.
\toolname generates more high-quality test cases by applying various text transformation methods (e.g., change tense) to the text in a failed test cases.
By doing so, \toolname can utilise a small text corpus to generate a large number of audio test cases, something which CrossASR++ is not capable of. 
In addition, \toolname implements more metrics to evaluate the performance of ASR systems from multiple perspectives. 
\toolname performs phonetic analysis on the identified failed test cases to identify the phonemes that ASR systems tend to transcribe incorrectly, providing useful information for developers to improve ASR systems.
The demonstration video of our tool is made online at \url{https://www.youtube.com/watch?v=DzVwfc3h9As}.
The implementation is available at \url{https://github.com/danielyuenhx/asdf-differential-testing}.

\end{abstract}

\section{Introduction}
The growing presence of Automated Speech Recognition (ASR) systems in modern society~\cite{8301638,robot_control,healthcare} motivates the need to properly test ASR systems. 
Researchers have proposed a series of methods to test various artificial intelligent systems (e.g., image classification~\cite{9825775,DeepHunter}, autonomous driving~\cite{acsac2022gong,deeproad}, etc) from various perspectives (e.g., fairness~\cite{9653830}, robustness~\cite{10.1145/3510003.3510146})
Recently, the rise of automated audio test case synthesis~\cite{ASRTest,asyrofi2021crossasr++,asrevolve,asrdebugger,10.1007/978-3-030-99429-7_14} has significantly reduced human involvement in the ASR testing process.
For example, CrossASR++~\cite{asyrofi2021crossasr++} is a tool that leverages Text-to-Speech (TTS) services to automatically generate audio files from texts and uses them to test ASR systems. 
Intuitively, the transcription produced by an ASR system should be equivalent to the text used to generate the audio.
Otherwise, a failed test case for an ASR system is uncovered.

Although CrossASR++ demonstrates the capability of uncovering failed test cases successfully, it still has a few limitations. 
First, it fails to make efficient use of the text corpus;
it requires taking as input a large amount of texts to find sufficient failed test cases.
The variation and volume of generated test cases depends solely on the quality and quantity of the provided text corpus. 
Such exhaustive selection employed in CrossASR++ is time-consuming and inefficient.
Second, CrossASR++ only reports the number of failed test cases uncovered, while more fine-grained information is needed to help developers to improve ASR systems.

To tackle the aforementioned limitations, we propose \textsc{Asdf}, a new \underline{A}utomated \underline{S}peech Recognition \underline{D}ifferential Testing \underline{F}ramework to test ASR systems. 
Our tool has the following features. 
It employs a \textit{text transformation module} and leverages known errors in ASR systems to synthesise multiple audio test cases from a single text.
After collecting a small initial set of failed test cases, \toolname transforms the failed texts using various text transformation methods to further generate more test cases.
For example, changing the tense of a sentence or substituting error-inducing terms with other words that have similar phonemes.
Our experiments show that utilising the text transformation module can boost the number of failed texts by an average of 22.3\%. Details of our experiment results can be found in our GitHub repository~\cite{asdf}.
\toolname also conducts phonetic analysis to identify the phonemes that are more challenging for ASR systems to transcribe. 
The phoneme information can provide useful information for developers to further improve ASR systems and their robustness~\cite{Fang2020}.

\section{Tool Design}

\begin{figure}[!t]
\centering
\includegraphics[width=9cm]{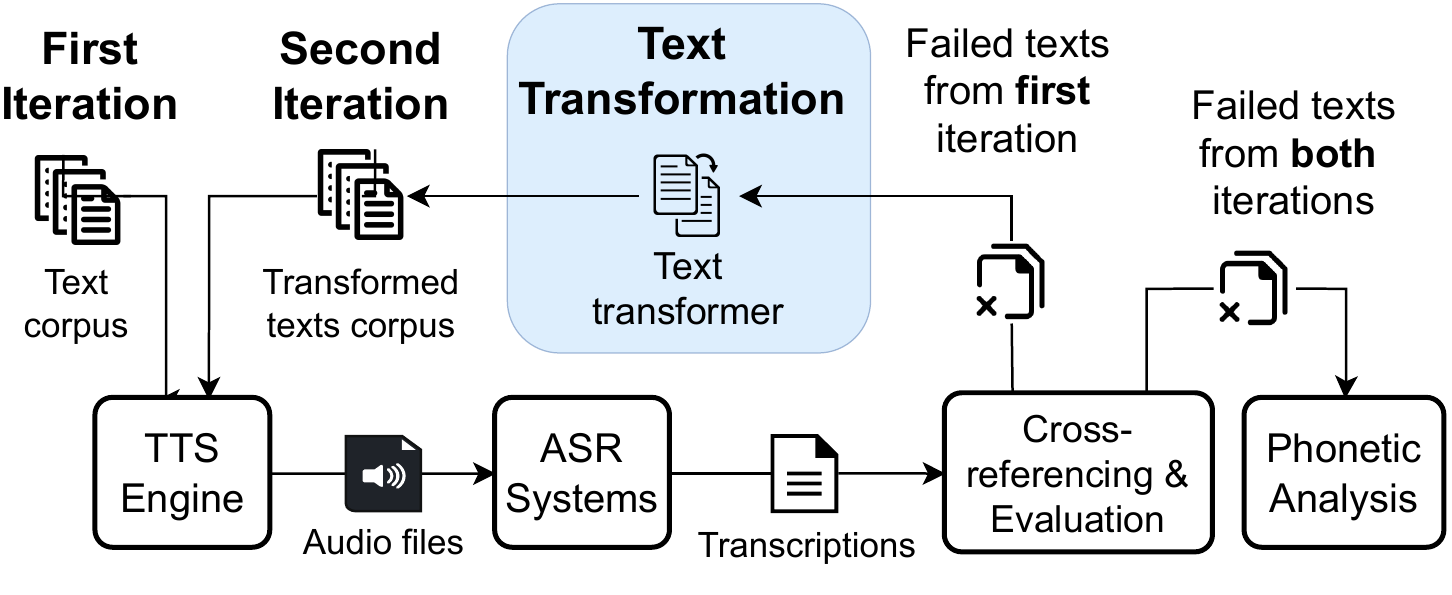}
\caption{The overview of the pipeline to generate test cases in \textsc{Asdf}.}
\label{fig:overview}
\end{figure}

\toolname extends the differential testing workflow used in CrossASR++~\cite{crossasr++}. 
Figure~\ref{fig:overview} illustrates the pipeline used by \toolname to generate test cases.
The pipeline involves two iterations. 
In the first iteration, each text in the corpus is processed as follows.
A Text-To-Speech (TTS) service is used to generate audio from the text, which is then fed into the ASR systems to obtain their transcriptions. 
Then, we perform cross-referencing by comparing the outputs from various ASR systems.
If there exists at least one system that can correctly transcribe the audio while a specific ASR system fails to do so, we say that a \textit{failed test case} is uncovered for this specific ASR system.
We refer interested readers to our previous paper~\cite{asyrofi2021crossasr++} for a more detailed description of this pipeline. 
\toolname can be easily configured to test multiple ASR systems. 
The installation, configuration, and usage instructions for \toolname are made publicly available in our GitHub repository~\cite{asdf}.

The failed test cases (i.e., original texts and their corresponding transcriptions) uncovered in the first iteration are collected and analyzed in the second iteration, which includes a text transformation module. 
In this iteration, a text transformer will mutate the texts of the failed test cases, effectively generating one or more transformed test cases to be added back to the corpus. This stage is crucial in extending the number of test cases available in the original corpus. 

Lastly, phonetic analysis is performed on the failed test cases from both iterations. The phonetics of the error-inducing terms found in the test cases will be analyzed to determine the most common phonemes that appear in the failed test cases. Analysis of highly occurring phonemes in error-inducing terms can be utilised to identify the phonemes that are more challenging for ASR systems to transcribe. These error-prone phonemes can be targeted for further improvements or research in the context of ASR transcription.

\section{Core Functionalities}
\toolname requires 5 inputs: 1) a text corpus, 2) the output directory of results, 3) the number of texts to be processed, 4) the ASR systems under test, and 5) a text transformation method. 
The input text corpus should be in the format of a \texttt{.txt} file. 
Each line of text in the file is converted into an audio file by gTTS~\cite{gtts}. 
Users can specify the number of texts to be processed before the \textit{Text Transformation Stage} begins. 
The ASR systems currently available in \toolname are DeepSpeech~\cite{deepspeech}, wav2letter~\cite{wav2letter}, and wav2vec2~\cite{wav2vec2}. 
Other ASR systems can be easily added and tested by inheriting the ASR abstract class, adhering to the interface rules~\cite{crossasr++}. 

Cross-referencing is performed amongst the selected ASR systems to ensure that the test cases are valid and can be determined by at least one ASR service. 
If specified, a text transformer transforms the original texts of the failed test cases found in the first iteration. This step generates new test cases to be appended to the original corpus, which is used to test ASR systems in the second iteration. 
Lastly, the test results are outputted to the path as specified. 

Currently, \toolname supports multiple transformation methods. 
The \textit{Homophone Transformation} method firstly identifies the homophone of the error-inducing term of the failed test case. 
Subsequently, a new example sentence that contains that homophone is obtained through the \texttt{WordHoard} Python library~\cite{wordhoard}. 
This new sentence will be used as the new test case. 
Details and examples of the other available text transformation methods such as Augmentation, Adjacent Deletion, Plurality Transformation, and Tense Transformation can be found in our replication package~\cite{asdf}. 

\section{Usage}
Ten metrics are used to evaluate the performance of the ASR systems. 
A bar chart is also plotted showing the phonemes and their frequency of appearance in error-inducing terms. 
A \texttt{.csv} file is produced, which can be used to perform analysis on the performance of each individual ASR system. Details of the metrics can be found in the GitHub repository~\cite{asdf}.

A failed text is defined as an audio file that is incorrectly transcribed by at least one ASR system. 
It is important to note that cross-referencing is used to filter the low-quality, unrealistic texts resulting from text transformation. 
A transformed text input must be transcribed correctly by at least one ASR service to be deemed valid (or determinable); otherwise, it is deemed as indeterminable and is discarded.

The percentage of transformed failed text is the ratio of failed transformed text inputs to the total transformed text inputs.
The higher the percentage of transformed failed text, the better the quality of the transformation method, as this indicates that its transformed texts are more challenging for ASR systems to transcribe correctly. 
A failed test case is defined as a specific text output from an individual ASR service that does not match its corresponding input text.  

The percentage of transformed failed cases is the ratio of transformed output from the ASR systems that does not match with its corresponding transformed input to the total number of transformed outputs from the ASR systems. The higher the percentage of failed transformed cases, the better the coverage of the transformed texts generated across different ASR systems. 

\section{Challenges and Future Work}
\toolname generates an audio test suite using a TTS library. 
This assumes that the generated audio is interchangeable with the real-world speech of humans, which may not be the case in real-life situations. There would still be differences between human speech and computer-generated audio, namely pronunciation, accent, and tone. As a result, if human speech were to be used instead of computer-generated audio, such inconsistencies may yield different results for our differential testing.
Furthermore, more sophisticated text transformation strategies can be incorporated into differential testing to test their effectiveness in revealing erroneous words. Such techniques include a change in grammar, a change in words, and a modification of sentence structure.

\balance

\bibliographystyle{IEEEtran}
\bibliography{bibliography}

\vspace{12pt}

\end{document}